\documentclass[12pt]{iopart} 
\pdfoutput=1
\usepackage{graphicx} 
\usepackage[caption=false]{subfig}
\usepackage{color}
\usepackage{float}

\begin{document} 

\title[]{Exploring the onset of collective 
motion in  self-organised trails of social organisms
}

\author{ E. Brigatti$^{\star}$ and A. Hern\'andez$^{\star}$$^{\ddag}$
}
\address{$^{\star}$ Instituto de F\'{\i}sica, Universidade Federal do Rio de Janeiro, 
Av. Athos da Silveira Ramos, 149,
Cidade Universitária, 21941-972, Rio de Janeiro, RJ, Brazil}
\address{$^{\ddag}$Instituto de Biocomputaci\'on y F\'{\i}sica de Sistemas Complejos (BIFI), Universidad de Zaragoza, Mariano Esquillor s/n, 50018 Zaragoza, Spain}

\ead{edgardo@if.ufrj.br}

\begin{abstract} 

We investigate the emergence of self-organised trails 
between two specific target areas
in collective motion of social organisms by means of
an agent-based model.
We present numerical evidences that 
an increase in the efficiency
of navigation 
in dependence of the colony size, exists. 
Moreover, the shift, from 
the
diffusive to the directed motion 
can be quantitatively characterised, 
identifying and measuring 
a well defined crossover point.
This point corresponds to the
minimal number of individuals
necessary for the onset of collective cooperation. 
Finally, by means of a finite-size scaling analysis, 
we describe its scaling behavior as a function of the environment size.
This last result can be of particular interest
for interpreting empirical observations 
or for the design of artificial swarms. 

\end{abstract} 

\vspace{0.8cm}

\maketitle

\section{Introduction}

Collective motion of social organisms 
is an elegant example of an emergent phenomenon
that can produce 
efficient behaviours 
based on a distributed cooperative cognition \cite{cognition}. 
Among the different structures that this phenomenon 
can generate \cite{swarm,ants1}, one of the more intriguing
is the self-organisation of trails, both in ants colonies \cite{ants2,ants3}, 
and in human communities \cite{pedestrian}.
Ant trail formation can be generated by means of the deposition 
of chemical pheromones
that enables indirect communication among ants through an environmental marking procedure.
By means of this mechanism, ants implement a reinforcement rule that
allows 
for selection of the shortest path to connect food source to nest location.
In this way, an adaptive behaviour  based only on local information and interaction is achieved.

This astonishing behaviour has motivated 
important technological applications.  One of the most notorious is a technique for 
general purpose optimisation \cite{dorigo}. 
A more recent one is the experimental implementation
of a navigation strategy for swarms of robots challenged to find a path between two target areas in an unknown environment \cite{robots}.
The solution of this practical problem 
opened new questions in relation to
the scalability 
of this approach to larger groups and larger environment size.
In other words, the characterisation of
the behaviour of this strategy of navigation in dependence of the group size
and its scaling in dependence of the environment size became 
a central topic of investigation.
An earlier result, related to the problem of the dependence on the community 
size, was obtained collecting field data from real ant colonies 
\cite{beekman}. 
This study shown a general increase in the number of ants walking to the feeder along the trail in relation to colony size and a very simplified 
mean-field model suggested that a minimum number of ants is required for an effective trail formation. 
Inspired by these previous results, the purpose of our work is to study and clearly characterise the nature of these specific aspects of trail formation 
by means of an accurate numerical analysis of the results produced by  a microscopic model which directly generates the trails.\\

Continuum microscopic models which are able to describe the self-organisation of 
ants trail formation are well known in the literature \cite{schweitzer,schweitzer2}.
These models are related to the general class of active Brownian particle models.
The motion of this random particles is determined by a
field which is directly influenced by the movement of the particles themselves.
This non-linear feedback, which operates between the particles and 
the generated field  at microscopic level, 
results in the self-organisation of the trails at the macroscopic level.
The use of this formalism allows the introduction of analytical approximations 
for achieving general results. 
For example, in \cite{agent2}, a mean-field approximation
determined the line that separates the system phase 
exhibiting a pure diffusion from the one where spatial structures 
of a general type can emerge.
Note that these studies can determine theoretically
the crossover line, but not the resulting patterns.
Otherwise, explicit solutions and simulations of
the process are obtained by means of the discretisation 
and numerical solution of the continuous model.
Another approach 
consider a discretised space and time 
where the motion of each individual 
is described by 
some transition rules. 
These agent based simulations   \cite{agent1,agent2,agent3} describe the same 
process of the continuum models, by means of 
these rules, which implement
the movements and the deposition of the pheromone. 

The use of agent based models is very important since
they allow for the understanding of the role of fluctuations and noise, 
as well as the limitations and validity of the continuous and the mean 
field descriptions. Indeed, the intrinsic 
stochasticity produced at the individual level
generates an internal noise which, in general, can cause 
impacting consequences \cite{edgardo1,toral}.
Moreover, as the central aim of our work is to understand what is the
minimum number of ants required for trail formation to become effective,
the description of the discrete nature of individuals 
is essential to characterise threshold and finite size effects. 
These effects can not be characterised by a continuum description where every small amount of the density of population is acceptable, even if unrealistically small \cite{edgardo2}. 
\\

In the following (Sec. 2), we introduce the details of the agent based model.
Even if it is inspired by previous works \cite{schweitzer2,agent2},
in order to implement an in deep numerical analysis of the phenomenon, 
and not just 
some specific examples of trails formation,
we consider a more simplified modelling approach.  In fact, the model counts on a single pheromone and its
dynamics depends on only two parameters. Trail formation is obtained based only on local information and interaction. 
In Sec. 3 we report the numerical analysis for the characterisation 
of the efficiency of navigation 
and for the quantitative description of the shift from the 
diffusive to the directed motion.
We would like to stress that the aim of the work is not to
identify a classical phase transition,
but rather we are interested in the scaling behaviour for finite-size 
systems. A discussion of these important points can be found 
at the end of the paper.

\section{The model}

The ant colony is composed by a population of $P$ individuals.
They can move on a regular 2-dimensional square 
lattice with $L \times L$ sites and periodic boundary conditions.
We choose $L$ odd, with the origin of the 
coordinate in the centre of the lattice.
The nest is located in (0,0) and the food source 
in (0,D), with $D=(L-1)/2$.
The time unit $t$ is the time interval between two updatings of the 
positions of all the individuals of the colony.

In the initial state each ant is located at the nest. 
An individual at site $(x,y)$ can move only on the top site ($x,y+1$) 
and on the right ($x+1,y$) or the left one ($x-1,y$). Steps towards
the bottom site are forbidden. In this way, individuals
effectively walk along paths where no loops are allowed.
This rule, in a simplified  form, takes into account the ants 
persistence to keep the direction of motion \cite{schweitzer2,orient},
reducing the probability of moving abruptly backward before reaching 
the specific goal. 
In fact, various pheromone-following ants can correct their walks 
by using environmental 
and even magnetic cues  \cite{nature}.  


We have also implemented a model where ants diffuse in all the four lattice
directions until they find the food source. In this case, results are equivalent, just a 
longer transition towards the quasi-stationary state is observed.

When an ant reaches the objective site, 
which represents the food source,
the possible directions of motion change, 
with all the movements allowed, except the
step towards the top site. 
Moreover, the ant starts to deposit,
in the new visited sites, an amount of pheromone $\phi=\exp[-\gamma(t-t_{act})]$.
Here $t_{act}$ is the time when the ant left the food, and 
$\gamma$ controls the critical time for an effective deposition.
This means that ants can effectively mark their trajectories
only for a limited time after they left the food.
When an ant reaches another time the nest,
$t_{act}=t$, making it effective in depositing pheromone
once more. 
This actualisation of $t_{act}$ is only implemented when
an ant reaches the food from the nest or the nest from the food.
It follows that ants which get lost are ineffective in depositing 
the pheromone. This function models
the tendency to deposit pheromone in association with specific 
stimuli and conditions \cite{schweitzer2,stimuli},
in this case the relative nearness of the nest or of the food,
instead of considering ants which deposit pheromone all the time
and in all the regions they visit.

At each time step all the deposited pheromone evaporates 
with a rate equal to $\epsilon$: 
$\phi_{t+1}(x,y)=(1-\epsilon)\times\phi_{t}(x,y)$.

Sites with a higher level of pheromone 
are more prone to be visited.
In fact, ants move to site ($x,y$) with 
probability $(1+\phi_{t}(x,y))/(S+3)$, where $S$
is the sum of all the pheromone presents in all the
neighbour sites and the other therms ensure that
if the pheromone is absent there is an equal probability
to move in the three possible directions. \\


\begin{table}
    \centering
{\small
    \begin{tabular}{|c|p{120px}|c|}
\hline
Parameter & Description \\
\hline
\hline
P & colony size \\
\hline
L & linear size of the system\\
\hline
$\gamma$ & pheromone effective deposition constant \\
\hline
$\epsilon$ & pheromone evaporation rate \\
\hline
\end{tabular}
}
\caption{
Parameters of the model. }
\label{table:params}
\end{table}

\section{Results and discussion}

\begin{figure}[h]
\begin{center}
\includegraphics[angle=0,width=0.8\textwidth]{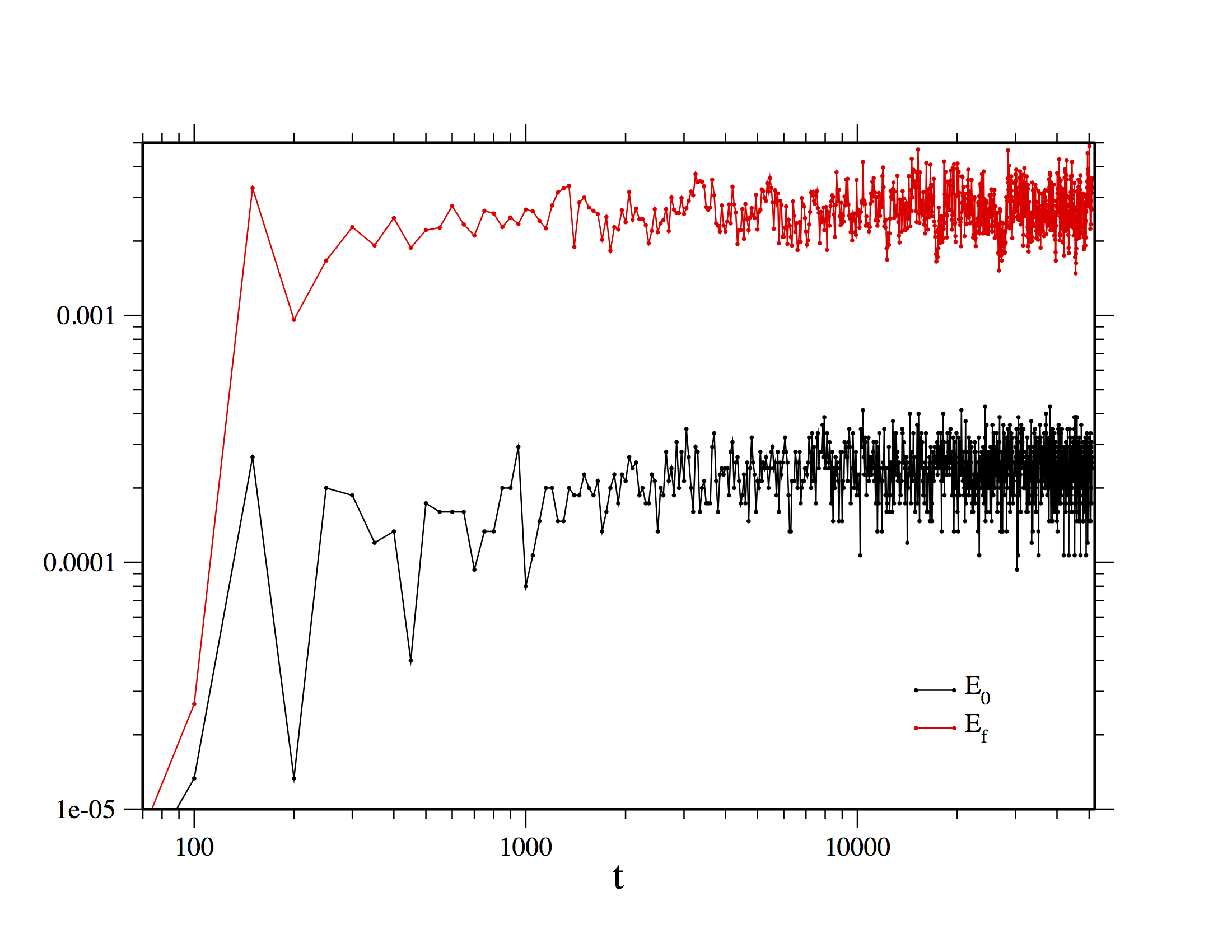}
\end{center}
\caption{\small
The time evolution of efficiency for a system with $L=41$, $\gamma=0.2$, 
$\epsilon=0.05$, and $P=1500$.  The nest is located in (0,0), the food in (0,20).
The points represent a time average over 50 time steps.
The lower curve represents the case where no pheromone is deposited ($E_0$).
  }
\label{Fig_dynamics}
\end{figure}

We run different simulations aiming at exploring the onset of the cooperative motion 
which allows the emergence of trails. 
The passage from a system where 
a diffusive behaviour is present, towards a system where a short path
is selected, can be easily monitored measuring the efficiency
of navigation between the two target areas.
This is achieved counting the number 
of ants which realise the trajectory from the nest to the food
in a time unit,   
normalised over the total population ($E_f$). 
The result is 
comparable if we measure the number 
of ants which realise the trajectory from the food to the nest.
In fact, after passing the transient time, there is
a rough symmetry in realising the two tasks.
In the case where no pheromone is deposited ($E_0$), 
the efficiency depends just on the value of $L$. 
This efficiency value describes a pure diffusive behaviour.
In contrast, if deposition is present, the efficiency is strongly dependent on all the 
parameters: $\gamma$, $\epsilon$, $P$ and $L$.
As it can be seen in Figure \ref{Fig_dynamics}, their dynamics are quite simple: after a fast transient the system reaches a quasi-stationary state where the value of $E_f$ and $E_0$ are maintained around a plateau.  
A good parameter 
for the description of the state of the system is $E=E_f/E_0$, 
which measures the gain in transportation efficiency 
for systems with self-organised trails. 
Organised states, where trail-based foraging emerged, 
present $E$ values clearly greater than the unit.

\begin{figure}[h]
\begin{center}
\includegraphics[angle=0,width=0.8\textwidth]{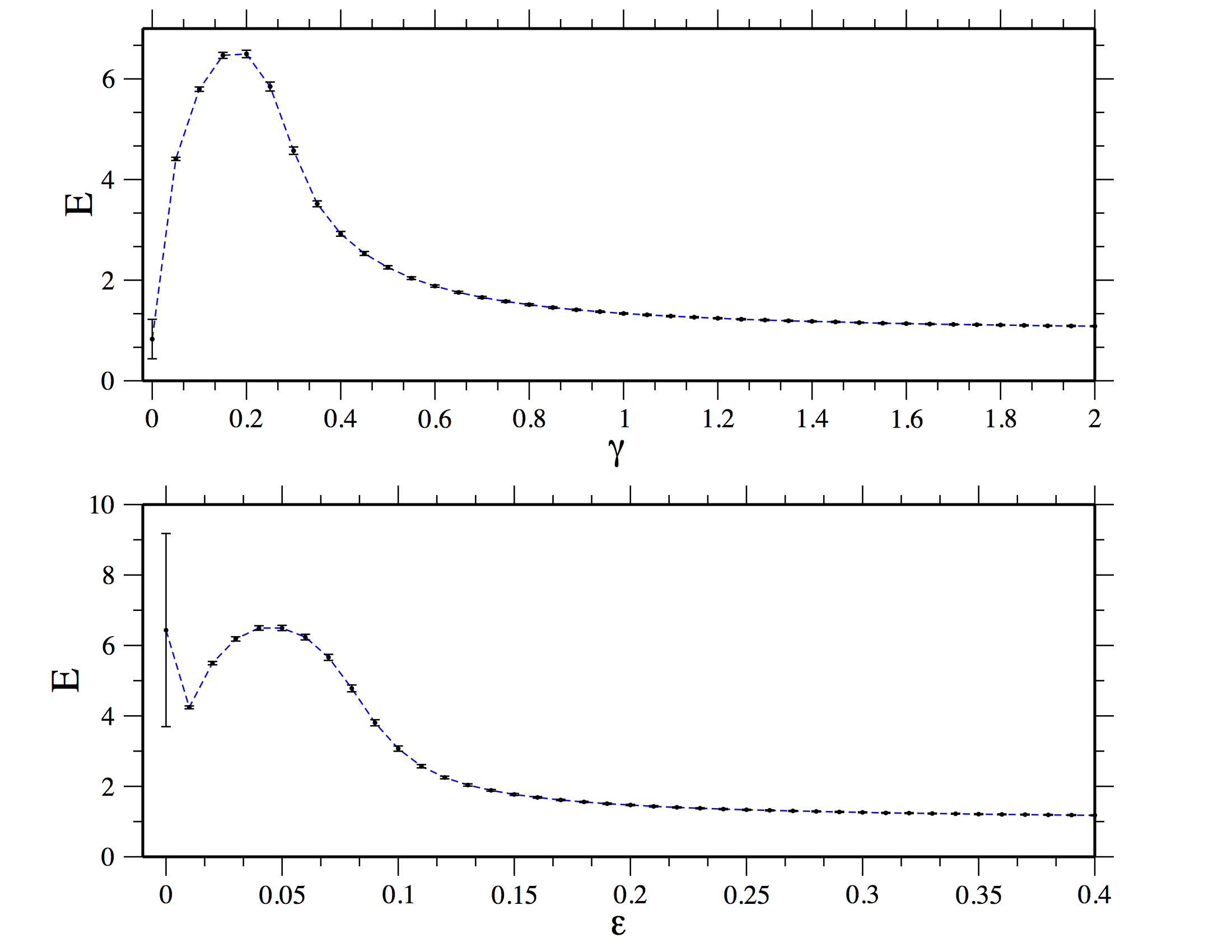}
\end{center}
\caption{\small
  Top: $E=E(\gamma)$, $\epsilon=0.05$.
  Bottom: $E=E(\epsilon)$, $\gamma=0.2$.
For all simulations  $L=17$, $P=100$ and the food is located in (0,8). 
Points are obtained from a temporal average over an interval of 50000 time steps 
and realising an ensemble average over 100 simulations.
Bars represent the value of the standard deviation of the ensemble average.   
}
\label{Fig_parameters}
\end{figure}

The dependence of $E$ as a function of  $\gamma$ and $\epsilon$ 
has a clear 
 behaviour.
In the top of Figure \ref{Fig_parameters} the behaviour of $E$ is depicted
for a fixed value of $\epsilon$. Changing the value of the 
pheromone effective deposition, $E$ grows from small values, when the deposition
is really fable (small $\gamma$), towards a maximal value for an optimal $\gamma$.
Then, it returns to smaller values, when the deposition remains 
active also for the lost ants, increasing the noise level in the reinforced paths.

In the bottom of 
figure \ref{Fig_parameters}, 
we can appreciate the behaviour of $E$
for a fixed value of $\gamma$. As for the previous case, a maximum value of $E$ exists in correspondence with an intermediate value of $\epsilon$. For larger value of evaporation the system, quite obviously, looses efficiency. 
Diminishing the value of $\epsilon$, the same behaviour is obtained.

Note that if the evaporation is absent, $E$ grows sensibly. 
This is followed by an impressive growth in the value of the variance of the ensemble average. 
In fact, for $\epsilon=0$, if the system selects an optimal path in the first period of the simulation, high levels of efficiency are registered, otherwise low levels are reached.
In this state the system is not effectively maximising its efficiency through a collective mechanism of exploration and signalisation, and efficiency strongly depends on the random configurations determined by the first paths.

Fixing the value of $\epsilon$ and $\gamma$, we 
turn to the most important part of our analysis. 
Our goal is to describe the 
onset of the cooperative motion 
that allows
for the emergence of trails. This corresponds to 
estimate the minimal number of ants
necessary for reaching the organised 
state.
For this reason, we must characterise 
the behaviour of the efficiency
in dependence of the colony size. 
Then, we analyse its scaling behaviour 
as a function of the environment size.

\begin{figure}[h]
\begin{center}
\includegraphics[angle=0,width=0.8\textwidth]{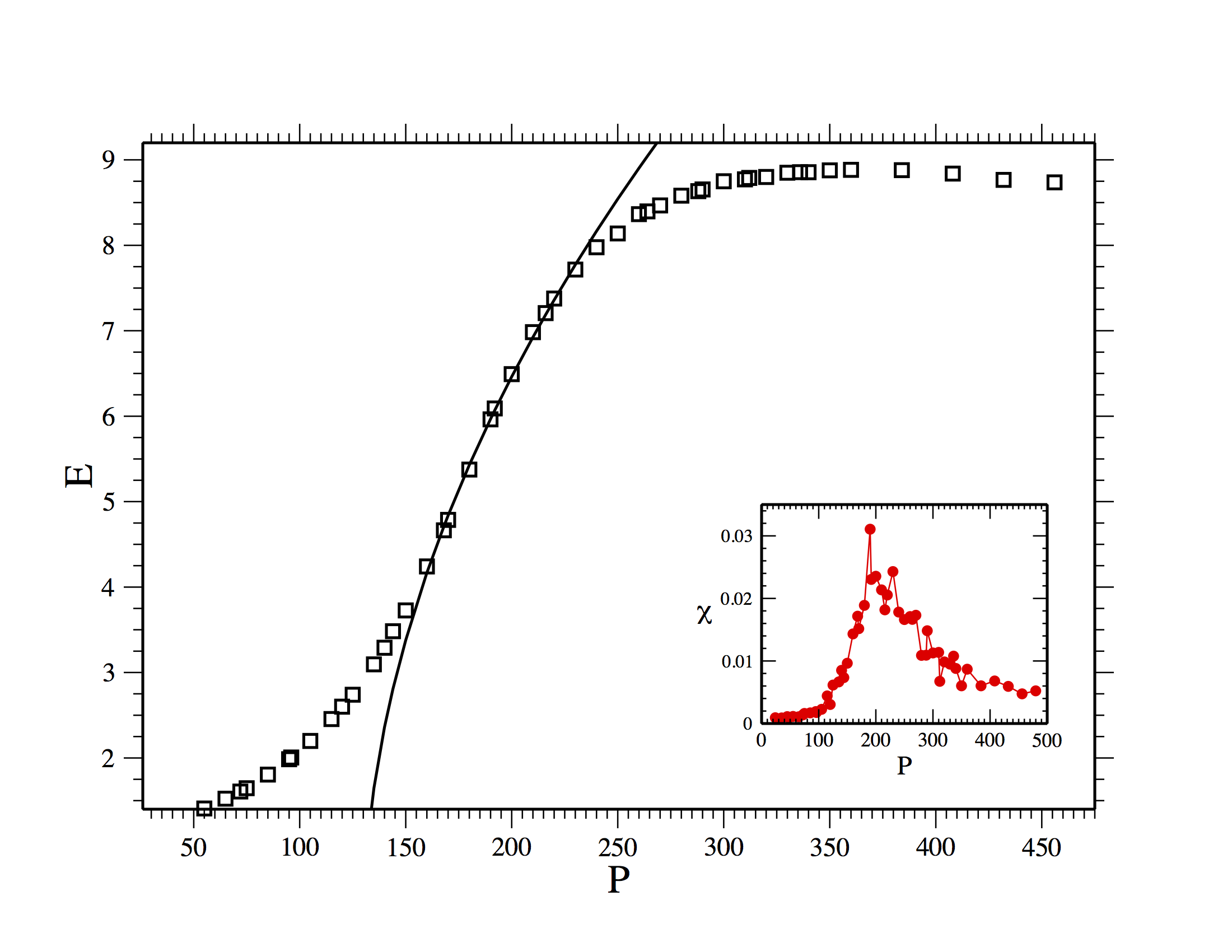}
\end{center}
\caption{\small
$E$ in dependence of $P$ for $L=25$, $\epsilon=0.05$, and $\gamma=0.2$.
The continuous line is the fitted power law: $y\propto (x-130)^{0.52}$.
In the inset, the behaviour of the variance $\chi$ is displayed.
Results have been averaged over 100 simulations.
}
\label{Fig_transit}
\end{figure}

Figure \ref{Fig_transit} shows the clear increase in the efficiency
of navigation between the two target areas
in relation to colony size $P$.
The depicted function has a typical sigmoidal shape. 
For small colony size a low value of $E$ is present, which corresponds
to a 
diffusive motion.
Increasing the colony size, a plateau
with a high value of $E$ is reached, which corresponds to the 
organised state. 
For higher values of $P$, $E$ starts to slowly decrease. 
This is probably due to the fact that very large populations
increase noise and generate saturation effects.
In general, our results are in accordance with the field study results 
presented in \cite{beekman}, which 
suggests
that a minimum number of ants is needed for enabling these trails.
To our knowledge, this is the first time this fact is clearly
reported by means of an agent-based microscopic model that describes 
the emergence of trails between two target areas.
For the different problem of unspecific pattern formation, 
a qualitative example was outlined in \cite{chialvo}.
In the same figure 
we show the behaviour of the variance $\chi$,
which can be obtained 
by evaluating:
$\chi=\langle E^{2}\rangle - \langle E \rangle^{2}$, 
where $\langle\;\;\rangle$ stands for the average over different simulations taken at the steady states. 
The existence of a clear peak 
of the variance 
suggests the location of a crossover point.
Indeed, the identification of the position of the maximum of this quantity
has been demonstrated to be a very robust approach for 
performing finite-size scaling analyses for classical equilibrium 
and for 
unconventional out-of-equilibrium systems
\cite{privman,edgardo3}.

For estimating the minimal number of ants
necessary to reach the 
organised
state ($P_m$) we 
also use another approach \cite{vicsek}. 
We note that 
the behaviour of $E$ is fairly
analogous to that of the order parameter of some 
equilibrium system, as
the region close to the 
point $P_m$ can be described by 
the approximation: $E\propto (P-P_m)^\alpha$
(see the continuous line in Figure \ref{Fig_transit}).
We determine $\alpha$ and $P_m$
by plotting $\ln{E}$ as a function of $\ln([P-P_m]/P_m)$
and using the value of $P_m$ for which the plot is the straightest
\cite{vicsek}. 
Hence, we can give a systematic estimation of $P_m$.
The existence of a $P_m$ is in accordance with the experimental
results of \cite{beekman}, which found that a 
minimum number of ants is needed for the emergence of trails. 
More general relations between
recruitment strategies and colony size \cite{planque} are in agreement with
these findings.
Indeed, species with small colony size predominantly use solitary foraging.
For increasing sizes other alternative methods are used \cite{alternative}, 
and  large colonies commonly use scent trails analogous to the ones
we model \cite{scent}.
 

\begin{figure}[h]
\begin{center}
\includegraphics[angle=0,width=0.8\textwidth]{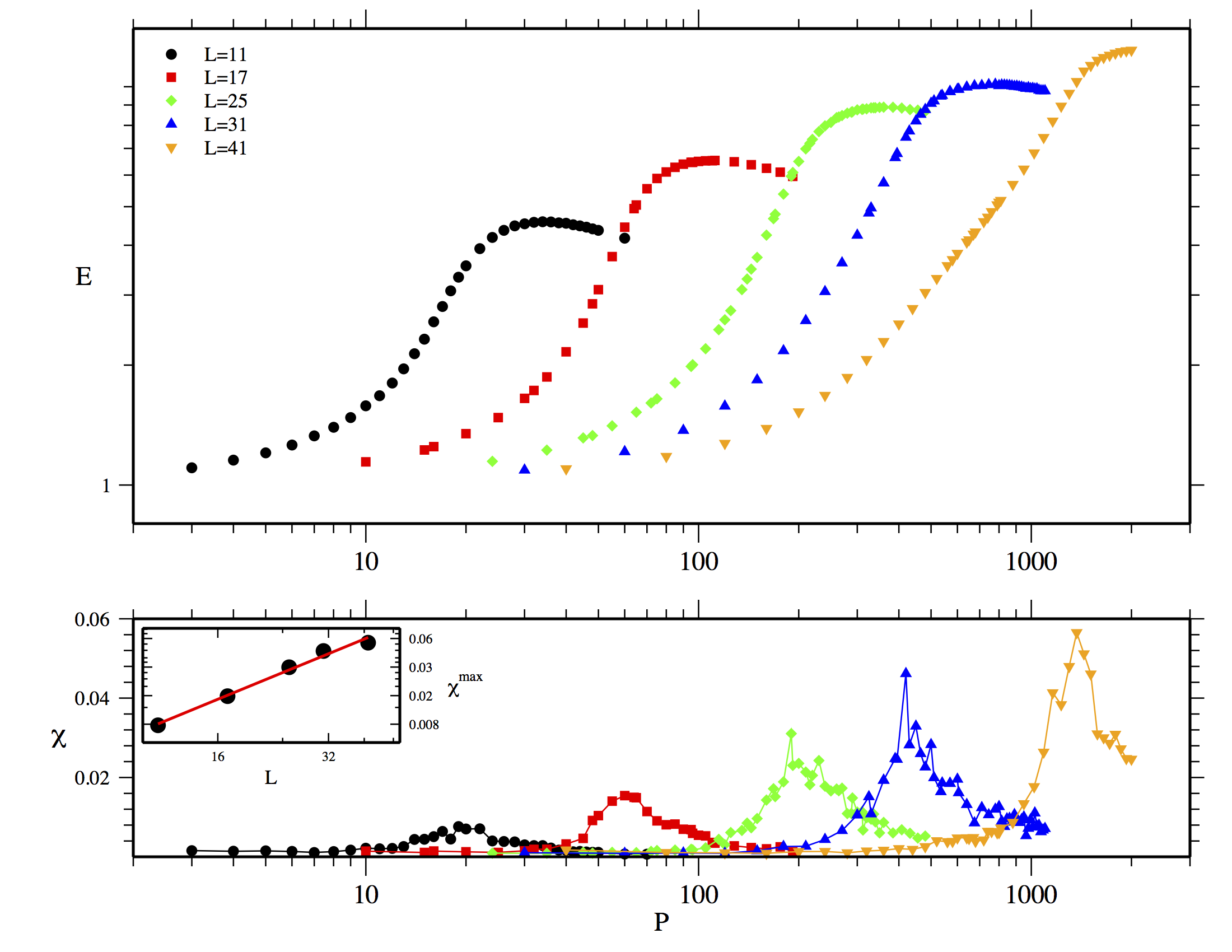}
\end{center}
\caption{\small
Top: $E$ in dependence of $P$ for different values of $L$.
Bottom: The variance $\chi$ in dependence of $P$ for different values of $L$. 
In the inset, the 
scaling of the maxima of $\chi$ in dependence of $L$. 
Interestingly, the variance peaks 
grow 
following a power-law (continuous line), a behaviour typical of classical phase transitions \cite{privman}.
Each point is averaged over 100 simulations; $\epsilon=0.05$ and $\gamma=0.2$.
}
\label{Fig_intellig1}
\end{figure}

\begin{figure}[h]
\begin{center}
\includegraphics[angle=0,width=0.8\textwidth]{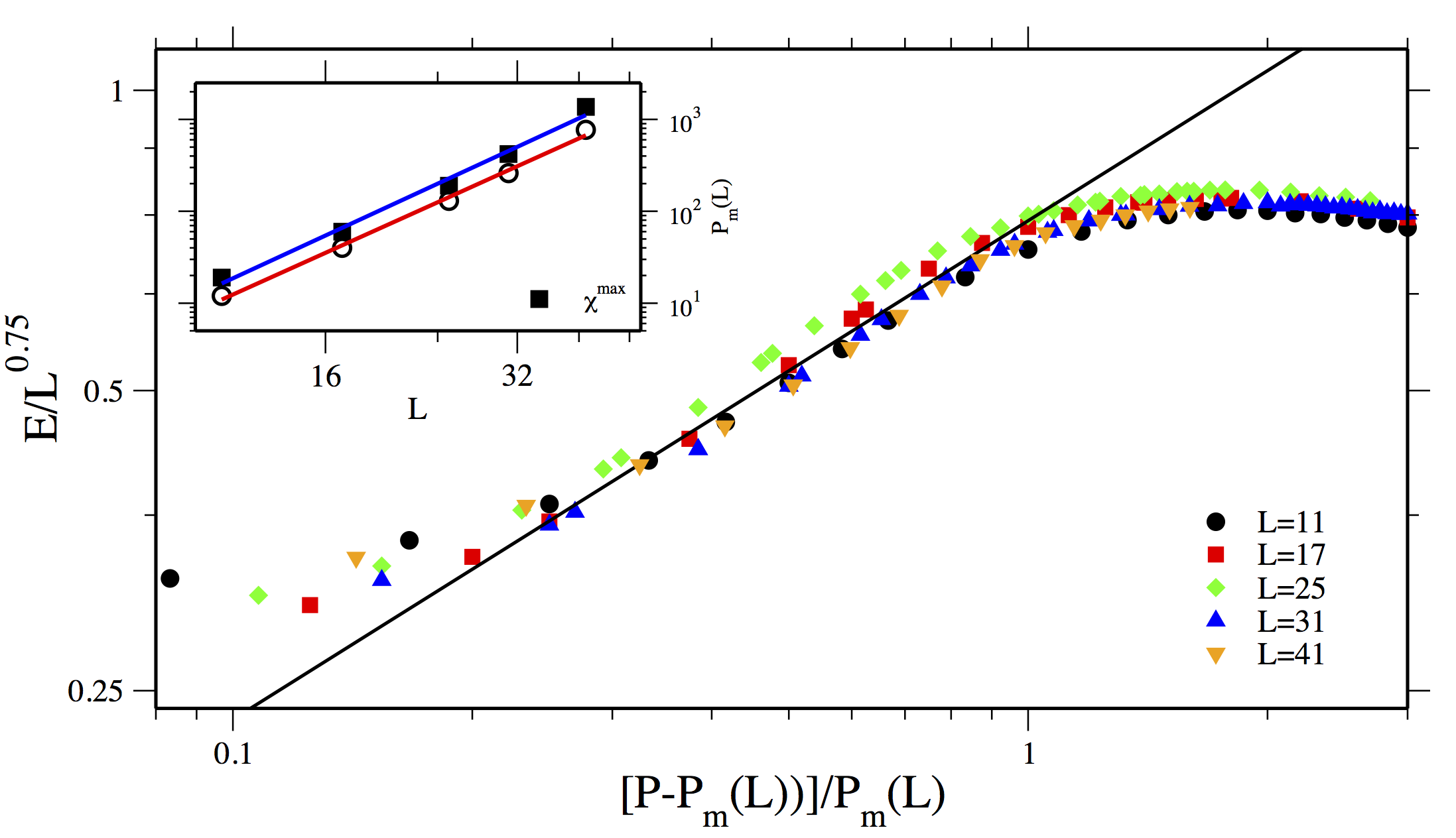}
\end{center}
\caption{\small
Rescaled logarithmic plot of $E$.
The continuous line has slope 0.5. Results have been averaged over 100 samples.
In the inset, finite-size transition points $P_m(L)$ measured from the maximum of the variance (squares) and 
from the shift constant of the power law description near the crossover area (circles).
The continuous lines represent the best fitting power-law functions. 
}
\label{Fig_scaling}
\end{figure}

Now we turn our attention to the description of the scaling
of these results 
in dependence of the environment size.
By changing the value of $L$ we rescale the entire system
with the selection of
$D=(L-1)/2$.
In Figure \ref{Fig_intellig1}, we can see the behaviour of $E$
and its variance $\chi$,
as a function of $P$, for different $L$ values.
As expected, for finite sizes, $P_m$ is $L$ dependent: $P_m(L)$.
As 
before, we 
estimate the different values of $P_m(L)$ 
from the shift constant of the power law description near the crossover area.
Moreover, for checking the consistency and robustness 
of this approach, we rescale the
data points using the rescaled parameter $[P-P_m(L)]/P_m(L)$.
Efficiency is rescaled using the relation $E/L^{0.75}$,
which was found looking at the scaling of the maxima
of $E$ for different L.
 As shown in detail in Figure \ref{Fig_scaling}, it is possible to obtain a 
 reasonable collapse of all the curves.
 Data roughly collapse 
 presenting  a common rate of vanishing close to $(P-P_{m}(L))^{0.5}$,
 strongly supporting the validity of this method. 
In fact, this result 
hardly can be considered a mere coincidence and it further legitimates
the use of the power-law approximation for determining $P_m$.
Using this procedure, the shift 
from 
the diffusive to the directed motion 
occurs at a critical value
that depends on the system size as: $P_{m} \propto L^{3.1\pm0.2}$. 
Note that the same conclusion can be obtained 
estimating the crossover points using the $P$ values where $\chi$ reaches 
the maximum. In this case the scaling has the form:
$P_{m} \propto L^{3.2\pm0.2}$.
This scaling indicates that, in the thermodynamic limit, the transition point goes to
infinite.  
Strictly speaking, this means that the system does not display a classical 
phase transition, which is rigorously defined at the thermodynamic limit in which the number of constituents 
tends to infinity.
However, it is also true that every finite system presents
a well defined transition point. 
Even if this point is not a genuine critical point, 
it has a clear physical meaning as it is the value of the $P$ 
parameter that identifies the shift from 
the diffusive to the directed motion, corresponding to the 
minimal number of individuals necessary to reach the 
organised state. 
Other models, where the transition is only observed for finite size systems, 
disappearing in the thermodynamic limit, and which present 
system size scaling, are well known in the literature \cite{toral,tessone}.

In general, when we transfer tools of statistical physics 
to problems of social or biological sciences, 
the population size is always  considerably smaller than the Avogadro number,
and so, not so large to justify the thermodynamic limit and its results.
In fact, we are interested in the behaviour of finite-size systems,
where important phenomena can appear in dependence of the number 
of individuals \cite{toral}. 
In particular, in our case, we expect that the relevant scaling
is limited to populations 
ranging from a dozen of agents, as for a swarm of mobile robots, reaching 
less than half a 
billion of individuals, as can happen in supercolony of some ant species \cite{higashi}.\\


This finite-size scaling results can be interpreted considering the density
of the colony population ($P/L^2$):
it is not sufficient to maintain a constant density
to obtain the trail formation. 
This fact highlights the non-linearity of the phenomenon and 
it poses important 
constraints for real robot implementations.
Moreover, our results can be explained
in term of a simple geometric scaling.
In fact, the minimal value of density, which generates a trail-based 
foraging, scales with the linear size 
of the environment,  which is proportional to the size of the trail.
Also the scaling of $E$ with $L^{0.75}$ points out
how the dynamics is far from merely depending 
on individual density.
Remembering that the definition of $E$ is already
normalised for the colony size, we can 
suppose that larger environments, postponing
the effects of noise and saturation, allow 
the active interaction of larger community of individuals,
which are able to maximise more efficiently
the navigation problem. \\


The reported behaviour of the finite scaling of the system is 
the most important result of our work.
In fact, in analogy with classical models of phase transition,
we can expect that the minimal number of individuals
necessary to reach the 
organised state 
can depends on the details of the model.
This would not be  the case for the scaling laws,
which could show some universal character, not 
depending on model details.
In this perspective, we hope that our simplified results,
perhaps refined using a model implementing
more specific conditions, could be successfully used for interpreting 
empirical observations of the scaling behaviour of ant colonies or of
artificial swarms built by cooperative mobile robots. 
 



\section*{References}

\begin{thebibliography}{99}

\bibitem{cognition}
Ofer Feinerman, and Amos Korman, Jour. Exp. Biol.,  {\bf 220}, 73 (2017).

\bibitem{swarm}
A. Okubo, 
Adv. Biophys., {\bf 22}, 1 (1986);
A. Cavagna et al., Proc. Natl. Acad. Sci. U.S.A., {\bf 107}, 11865 (2010);
S. Viscido, M. Miller and D. S. Wethey, J. Theor. Biol., {\bf 217}, 183 (2002);
 Calv\~ao A.M., Brigatti E.
 PLoS ONE {\bf 9}, e94221 (2014);
 T. Vicsek, A. Zafeiris, Physics Reports {\bf 517}, 71 (2012).

\bibitem{ants1}
Guy Theraulaz, Jacques Gautrais, Scott Camazine and Jean-Louis Deneubourg
Phil. Trans. R. Soc. Lond. A  {\bf 361}, 1263 (2003).

\bibitem{pedestrian}
Dirk Helbing, Joachim Keltsch and P\'eter Moln\'ar,
Nature, {\bf 388},  47 (1997).

\bibitem{ants2}
E.O. Wilson, 
Anim. Behav. {\bf 10}, 134 (1962);
A. Dussutour, V. Fourcassie, D. Helbing and J.L. Deneubourg, 
Nature {\bf 248}, 70 (2004);
A. Dussutour, J.L. Deneubourg and V. Fourcassie, 
Journal of Experimental Biology {\bf 208}, 2903 (2005).

\bibitem{ants3}
Fonio et al., eLife {\bf 5}, e20185 (2016).

\bibitem{dorigo}
M. Dorigo, M. Birattari, and T. Stutzle
IEEE Comput. Intell. Mag. 28 (2006).

\bibitem{robots}
Valerio Sperati, Vito Trianni and, Stefano Nolfi,
Swarm Intell  {\bf 5}, 97 (2011)

\bibitem{beekman}
Madeleine Beekman, David J. T. Sumpter, and Francis L. W. Ratnieks,
PNAS  {\bf 98}, 9703 (2001).

\bibitem{schweitzer}
M. M. Millonas, J. Theor. Biol. 159, 529 (1992);
F. Schweitzer, K. Lao, and F. Family, BioSystems {\bf 41}, 153 (1997);

\bibitem{schweitzer2}
D. Helbing, F. Schweitzer, J. Keltsch, P. Molnar,  
Phys Rev E {\bf 56}, 2527 (1997).

\bibitem{agent1}
Deneubourg, J.L., Goss, S., Franks, N. and Pasteels, J.M., 
J. Insect Behav. {\bf 2}, (1989).
\bibitem{agent2} Erik M. Rauch, Mark M. Millonas and Dante R. Chialvo
Phys. Lett. A {\bf 207}, 185, (1995).
\bibitem{agent3} I.D. Couzin, and N.R. Franks,
Proc. R. Soc. Lond. B, 02PB0606 (2002).

\bibitem{edgardo1}
E. Brigatti,V. Schw\"ammle, and Minos A. Neto,
Phys. Rev. E, {\bf 77}, 021914 (2008).

\bibitem{toral}
R. Toral and C. J. Tessone, Commun. Comput. Phys., {\bf 2}, 177 (2007).


\bibitem{edgardo2}
E. Brigatti, M. Oliva, M. N\'u\~nez-L\'opez, R. Oliveros-Ramos and J. Benavides
EPL, {\bf 88}, 68002 (2009); E. Brigatti, M. N\'u\~nez-L\'opez, and M. Oliva
Eur. Phys. J. B {\bf 81}, 321 (2011).	

\bibitem{orient}
W. Alt, J. Math. Biol. {\bf 9}, 147 (1980).

\bibitem{nature}
D. E. Jackson, M. Holcombe, and F.L.W. Ratnieks, 
Nature, {\bf 432}, 907 (2004).

\bibitem{stimuli}
T.J. Czaczkes, C. Gr\"uter, and F.L.W. Ratnieks,  
J R Soc Interface {\bf10}, 20121009 (2013);
T.J. Czaczkes, and J. Heinze,  
Proc. R. Soc. B {\bf 282}, 20150679  (2015).


\bibitem{privman}
V. Privman, Finite Size Scaling and Numerical Simulations of
Statistical Systems (World Scientific, Singapore, 1990).

\bibitem{edgardo3}
R. Dickman and J. Kamphorst Leal da Silva, Phys. Rev. E 58, 4266 (1998);
Konstantin Klemm, V\'\i ctor M. Egu\'\i luz, Ra\'ul Toral, and Maxi San Miguel
Phys. Rev. E {\bf 67}, 026120 (2003); N. Crokidakis and E. Brigatti, J. Stat. Mech. P01019 (2015);
E. Brigatti and A. Hern\'andez, 
Phys. Rev. E {\bf 94}, 052308 (2016).

\bibitem{chialvo}
D. R. Chialvo, New Ideas Psych. {\bf 26}, 158 (2008).

\bibitem{vicsek}
T. Vicsek, et al. Phys. Rev. Lett., {\bf 75}, 1226 (1995).

\bibitem{planque}
Planqu\'e R, Van den Berg JB, Franks NR,  
PLoS ONE {\bf 5}, e11664 (2010).

\bibitem{alternative}
Holldobler B, Wilson EO  The Ants. Cambridge, Massachussets: Belknap, Harvard (1990).

\bibitem{scent}
Beckers R, Deneubourg JL, Goss S  
J Insect Behavior {\bf 6} 751 (1993);
Franks NR, Gomez N, Goss S, Deneubourg JL 
J Insect Behav {\bf 4} 583 (1991);
Franks NR, Sendova-Franks AB, Simmons J, Mogie M, 
Proc Roy Soc London B {\bf 266} 1697 (1999) .

\bibitem{tessone}
C. J. Tessone, R. Toral, P. Amengual, H. S. Wio, M. San Miguel, Eur. Phys. J. B, {\bf 39}, 535,(2004);
K. Klemm, V. M. Eguiluz, R. Toral, M. San Miguel, Phys. Rev. E {\bf 67}, 045101R (2003);
Carlos P. Herrero, Phys. Rev. E {\bf 69}, 067109  (2004).


\bibitem{higashi}
S. Higashi and, K. Yamauchi, 
Japanese Journal of Ecology, {\bf 29}, 257 (1979).




\end{thebibliography}
\end{document}